\documentclass[twocolumn,amsmath,amssymb,aps,notitlepage]{revtex4-1}
\usepackage[utf8]{inputenc}
\usepackage{bm}
\usepackage{comment}
\usepackage{graphicx}
\usepackage{dcolumn}
\usepackage{braket}
\usepackage[caption=false]{subfig}
\usepackage{url}
\usepackage{subfiles}
\usepackage{color}

\newcommand{\beq}{\begin{equation}}
\newcommand{\eeq}{\end{equation}}
\newcommand{\beqa}{\begin{eqnarray}}
\newcommand{\eeqa}{\end{eqnarray}}
\newcommand{\bem}{\begin{math}}
\newcommand{\eem}{\end{math}}

\newcommand{\bfr}{{\bf r}}

\usepackage{xcolor,soul}

\def\tl#1{\textcolor{black}{#1}}
\def\sc#1{\textcolor{black}{#1}}

\setstcolor{blue}

\begin{document}
 
\title{An equation of state for active matter 
%: Non-equilibrium thermodynamic averages in 2D active brownian particles
  }
\author{Samuel Cameron, Majid Mosayebi, Rachel Bennett and Tanniemola B. Liverpool}
\affiliation{%
  University of Bristol}%
\date{\today}

\begin{abstract}
\tl{We characterise the steady states 
of a suspension of two-dimensional active brownian particles (ABPs). 
  \sc{W}e calculate 
  the steady-state probability distribution 
   to lowest order in Peclet number.  We show that macroscopic quantities can be calculated in analogous way to equilibrium systems using this probability distribution.
  We then \sc{derive} expressions for the macroscopic pressure \sc{and position-orientation correlation functions}. We check our results by direct comparison with extensive numerical simulations. A key finding is the importance of many-body effective interactions even at very low densities\sc{.}
  }
\end{abstract}

\maketitle

Statistical mechanics tells us the probability that
a system is in a certain state, as long as \tl{it} is
at equilibrium~\cite{Plischke2005}. For instance, the degrees of freedom for any
system in the canonical ensemble are sampled from the Gibbs-Boltzmann
distribution, $\mathbb{P}_{eq} \propto e^{-U/(k_BT)}$ (where $U$ is the system's energy).
$\mathbb{P}_{eq}$ can then be used to determine macroscopic system properties,
effectively reducing a large proportion of equilibrium statistical
mechanics to a difficult exercise in integration.
  \tl{In contrast, the steady-state
probability distribution of a non-equilibrium  system in a non-equilibrium steady-state (NESS), $\mathbb{P}_{ss}$,  is
not known a-priori \cite{risken1996fokker}.} In fact, a NESS need not always exist. However, if a NESS exists and $\mathbb{P}_{ss}$ can be
determined, one should in principle be able to apply similar techniques to equilibrium statistical mechanics
%(e.g. calculation of partition functions by integration) 
on non-equilibrium systems in a NESS.
Herein lies the motivation of this letter: calculating
macroscopic properties of non-equilibrium systems in a NESS using methods
analogous to those of equilibrium statistical mechanics.

A popular subset of non-equilibrium phenomena, which we restrict our attention to here, are ``active'' matter systems
\cite{Gompper_2020,RevModPhys.85.1143}, which never reach an
equilibrium state due to the presence of internal (bulk) energy sources.
Early studies of these active systems were motivated by 
biological processes on a wide array of length scales
(from e.g. fluctuations of membranes \cite{Prost_1996}
to flocking birds \cite{PhysRevLett.75.1226,TonerTu95,TONER2005170}), but has increasingly found application in synthetic man-made systems~\cite{Howse2007a,Bricard2013}. 

The active system we study here, known as Active Brownian Particles (ABPs)~\cite{PhysRevLett.108.235702,Bialk__2013,PhysRevLett.114.198301,PhysRevLett.117.038103,Bickmann_2020},
 is a suspension
of spherical colloids in a solvent, which have a mechanism of
self-propulsion driving them out of equilibrium~\cite{Golestanian2005}. 
%Typically, this self propulsion is facilitated by chemically coating one half of each
% colloid with a substance that reacts with the solvent (known as Janus particles), though % other mechanisms (e.g. involving external fields) also provide self-propulsion. 
ABPs are widely studied due to the relative simplicity of their equations of motion, while capturing the essentials of active matter systems.

Typically theoretical descriptions of active systems (such as ABPs) approximate the many-particle dynamics by an effective one-particle distribution function. For some macroscopic quantities, these single-particle distribution functions give qualitatively similar behaviour to explicit particle based simulations~\cite{Bialk__2013,PhysRevLett.112.218304}. However quantitative comparison is only possible using (often many) phenomenological fitting parameters. 
%which must be fit to data.
Other macroscopic quantities (e.g. the pressure of an interacting active gas) which depend on particle correlations, are simply impossible to obtain using these approaches.

Theoretical frameworks which take account of the many-particle correlations, on the other hand, look 
%discouragingly 
formidably difficult. % if not impossible. 
However recently, a generic approach to fluctuating dynamical systems highlighting the role of non-vanishing
currents in non-equilibrium steady-states (NESS) was introduced, where the relaxation dynamics towards the NESS plays a key role~\cite{PhysRevE.101.042107}. In particular, \cite{PhysRevE.101.042107} introduces
the notion of ``typical trajectories'', which we use here to construct  an ansatz for $\mathbb{P}_{ss}$ amenable to analytical solution.

This letter has three main results. The first result is an
\tl{explicit expression for the steady-state probability
distribution of ABPs, $\mathbb{P}_{ss}$  in $2D$. The second result is an effective
interaction potential for ABPs which is independent of particle
orientation, but depends on many-body (as opposed to pair-wise)
interactions. The third result uses  $\mathbb{P}_{ss}$} to obtain an equation of state, an expression for the active brownian
swim pressure \cite{C5SM01412C} at low Peclet number. \tl{The van der Waals equation is a landmark in statistical mechanics generalising the ideal gas law to an equation of state for realistic gases, expressing the pressure as an expansion in density~\cite{Plischke2005}.  Here we obtain an equivalent for active matter by obtaining an equation of state that is an expansion in {\em both} density and activity (here encoded in Peclet number).} We also compute
another macroscopic average, which probes local correlations between
inter-particle position and orientation.
Comparison of our calculated swim pressure and the local
correlation function to direct simulations yields good
agreement between the two for a range
of system densities. 
%The results presented here distinguish themselves from previous work on non-equilibrium steady-states in active matter
% both in the calculation of $\mathbb{P}_{ss}$ for interacting ABPs, and its connection to macroscopic observables.

We consider a collection of $N$ active brownian particles on a plane of area $A=L^2$ with periodic boundary conditions (BCs). The $3N$
degrees of freedom satisfy the system of stochastic differential equations
for their positions on the plane, $\bm{r}_i$, \tl{and their directions $\bm{u}_i  =(\cos\theta_i,\sin\theta_i)$, of self-propulsion}
%\footnote{A more broad definition of active brownian particles is also
%  found in the literature \cite{Romanczuk2012}, but for our purposes ABPs
%  are defined via the specific equations of motion in eqn. (\ref{eqs:eoms}).}
\begin{subequations}\label{eqs:eoms}
  \begin{align}
    \gamma d\bm{r}_i (t) &= (-\bm{\nabla}_{{\bm{r}_i}}U
    \label{eq:eom_position}
    +\tilde{f_P}\bm{u}_i)dt +\gamma \sqrt{2D_t} d\bm{W}_{\bm{r}_i}(t),\\
    d\theta_i (t) &= \sqrt{2 \tilde{D}_r}d W_{\theta_i}(t) \quad , \quad
    \label{eq:eom_theta}
    \end{align}
\end{subequations}
where $W_\alpha$ are as usual Wiener processes.
To be concrete, the interaction potential
$U$ is the sum of pair-wise (purely repulsive) Weeks-Chandler-Anderson
(WCA) potentials $\mathcal{V}(r_{ij})$ with energy scale $\epsilon$.
Length is measured in units of the interaction
length $\sigma$, time in units of $\sigma^2/D_t$, and
energy in units of $D_t\gamma$, where $D_t$ and
$\gamma$ are the translational diffusion and drag,
respectively. The remaining parameters in our system
quantify the strength of the active force relative to fluctuations,
$f_P\equiv\tilde{f}_P\sigma/D_t\gamma$,
the relative magnitude of rotational diffusion,
$D_r\equiv\tilde{D}_r\sigma^2/D_t$, and the
density of particles in our system, $\rho\equiv N\sigma^2/A$.
Our default parameters in this letter are $\epsilon/(D_t\gamma)=1$
and $D_r=3$ (the latter corresponding to a no-slip boundary
condition between the particles and the solvent in
equilibrium systems).

\begin{figure}[h]
  \includegraphics[width=0.5\textwidth]{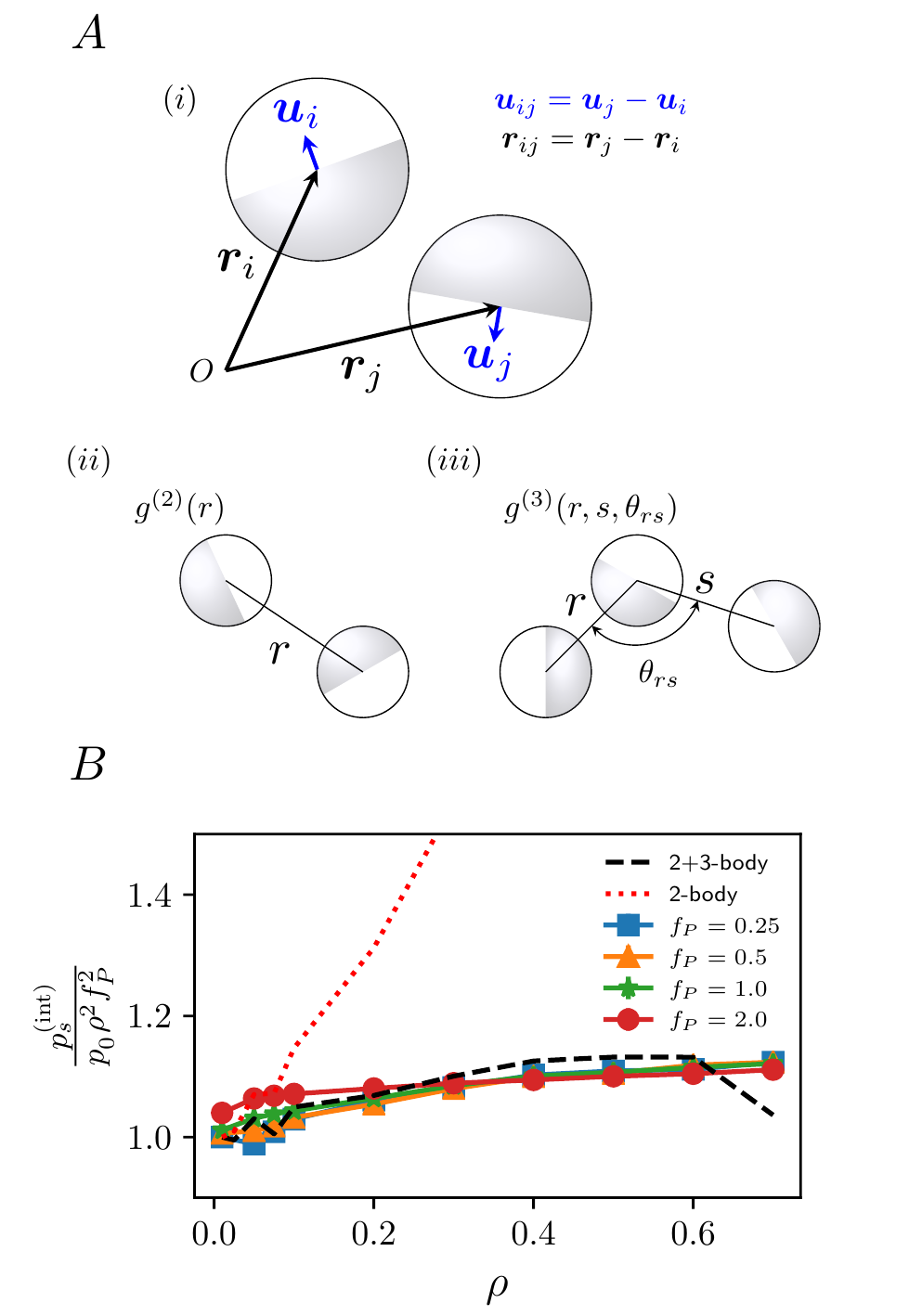}
  \caption{(A) Definitions of $\bm{u}_{ij}$ and $\bm{r}_{ij}$. Schematic of two and three body interactions. (B) Interacting swim pressure $p_s$ scaled by $\rho^2f_P^2 p_0$, where $\rho$ is
    density of ABPs, $f_P$ is the self-propulsion force and \tl{$p_0^{\mathrm{th}}$ is the swim pressure  at $\rho=0.01$ for the theory curve
      and $p_0^{\mathrm{sim}} = {p^{\mathrm{int}}_s \over \rho^2 f_p^2} (f_P=0.25,\rho=0.01)$
      for the all experimental (simulation) data}. Filled markers correspond
    to simulation data with
    self-propulsion values of $f_P=0.25$, $f_P=0.50$, \tl{$f_P=1.0$,} and $f_P=2.0$, respectively.
    The black solid line is our theoretical prediction. The red dashed line 
     excludes three-body correlations.}
     \label{fig:1}
\end{figure}

The eqns. (\ref{eqs:eoms})  are equivalent to a Fokker-Planck equation for the probability density, $\mathbb{P}_{ss}(\vec{X},t)$ of degrees of freedom
$\vec{X} = \left(\bm{r}_1,\theta_1,\bm{r}_2,\theta_2,\cdots,\bm{r}_N,\theta_N\right)$, 
%(now independent variables),
\begin{equation}\label{eq:FokkerPlanck}
  \partial_t\mathbb{P}\! = \!\!\sum_i\!\big(\partial_{\bm{r}_i}\cdot
  (\mathbb{P}\partial_{\bm{r}_i}U)
  -f_P\bm{u}_i\cdot\partial_{\bm{r}_i}\mathbb{P}
  +\partial_{\bm{r}_i}^2\mathbb{P}+D_r\partial_{\theta_i}^2\mathbb{P}\big)
\end{equation}
subject to periodic boundary conditions.
The steady-state probability density
$\mathbb{P}_{ss}(\vec{X})$ is a
solution of ($\partial_t\mathbb{P}_{ss}=0$). We
define a new function $\zeta(\vec{X})$ via 
\begin{math}\label{eq:zetadef}
  \mathbb{P}_{ss}(\vec{X}) \propto e^{-U(\vec{X}) - f_P\zeta(\vec{X})} \; , \; 
\end{math}
so that the equilibrium distribution is recovered when $f_P=0$.
 Inserting this 
into this Fokker-Planck equation maps the linear partial differential equation
(PDE) for $\mathbb{P}_{ss}(\vec{X})$ to a non-linear PDE for $\zeta(\vec{X})$.
This is helpful mathematically because $\zeta$ is less constrained than $\mathbb{P}$ and easier to approximate.
%However 
\tl{To calculate $\zeta$, we make some assumptions about its form.}
Symmetry under exchange of particles, translational and rotational invariance 
%and hence no macroscopic orientational order  
imply 
\begin{equation}\label{eq:zetatwobody}
  \zeta(\vec{X}) = 1/2\sum_{i,j\neq i}\bm{u}_{ij}\cdot\bm{r}_{ij}w(r_{ij}) + \mbox{h.o.t.} \quad ,
\end{equation}
where $r_{ij}=|\bm{r}_{ij}|,\bm{r}_{ij}=\bm{r}_j-\bm{r}_i$, $\bm{u}_{ij}=\bm{u}_j-\bm{u}_i$ and $w(r)$ is a scalar function (see Figure \ref{fig:1}A). \tl{We neglect terms that are nonlinear in orientation, assuming local alignment interactions are weak}. 

Finally, keeping terms to leading order in $f_P$, one arrives at  the ordinary differential equation (ODE) for $w(r)$
\begin{equation}
  w''(r)+ \Omega_1(r) w'(r)
  - \Omega_2(r) w(r)
  = \frac{\mathcal{V}'(r)}{2r} \; , 
\end{equation}
\sc{where
\begin{subequations}
  \begin{align}
    \Omega_1(r,\xi)&=3/r-\mathcal{V}'(r) - \xi,\\
    \Omega_2(r,\xi)&=D_r/2+\mathcal{V}'(r)/r - \xi/r
  \end{align}
\end{subequations}}
%\fx{\st{where $\ds \Omega_1=\bigg(\frac{3}{r}-\mathcal{V}'(r) - \xi\bigg)$,$\ds \Omega_2=\bigg(\frac{D_r}{2}+\frac{\mathcal{V}'(r)}{r} - \frac{\xi}{r}\bigg)$}} 
and $\xi$ fluctuates around zero, %$\left<{\xi}\right>=0$ 
    hence we set $\xi=0$ (see Methods).
\iffalse 
with
\beqa\left< \xi^2\right> &=& \frac12  \left\{ \rho \int d^2 r \,   g_2(\bfr)  \left(\mathcal{V}' (r_{})\right)^2 \right . \nonumber \\ && \left . + \rho^2 \int d^2 r \, d^2 r' \,  \hat\bfr \cdot \hat \bfr' g_3(\bfr,\bfr')  \left. \mathcal{V}' (r_{}) \mathcal{V}' (r'_{})\right. \right\}\; . \eeqa 
\fi
%The ODE is subject to the boundary conditions $w(\infty)=0$ and
This ODE is solvable using standard techniques
\cite{2020SciPy-NMeth,DORMAND198019,Lawrence1986SomePR}. Thus, the problem
of solving a $3N$ variable PDE for $\mathbb{P}_{ss}(\vec{X})$ in eqn
\ref{eq:FokkerPlanck} has been reduced to solving a single ODE. 
\tl{To solve for $w(r)$ (and hence $\zeta(\vec{X})$), we need two BC which we implement approximately as follows.
%We now make a couple of approximations to calculate $\zeta$.
First, we note that to be consistent with periodic BC's the range of $w(r)$ (like that of $\mathcal{V}(r)$) must be less than $L/2$, i.e. $w(r)=0, r> r_c$ with $r_c< L/2$~\cite{Frenkel2002}.
%, otherwise there will be interactions with images that must be taken account of.
As we consider large systems, $L \gg 2^{1/6}$, we relax this and simply require that $w (r) \to 0$ as $r \to \infty$.
%and check that $w(r)$ is very small at lengthscales comparable  $L/2$.
Next, we note that the typical
trajectories of a system in steady-state will generate (spatial) probability density
currents $\bm{J}_{i} %= (\bm{J}_1,\cdots,\bm{J}_N)
= \mathbb{P}_{ss}\bm{V}_{i}$ \cite{PhysRevE.101.042107}
with\sc{
\begin{equation}\label{eq:v_i}
  \bm{V}_i = f_P(\bm{u}_i + {\partial}_{\bm{r}_i}\zeta(\vec{X})) , \;   i \in \{ 1 ,\cdots, N\}.
\end{equation}}
%\fx{\st{
%\begin{math}\label{eq:v_i}
%  $\bm{V}_i = f_P(\bm{u}_i + {\partial}_{\bm{r}_i}\zeta(\vec{X})) , \;   i \in \{ 1 ,\cdots, N\}.$
    %\end{math} 
    %}}
These currents depend on $\mathbb{P}_{ss}$ and specify a family of trajectories, i.e. $\bm{V}_i$. This implies that we can reverse this logic and use the typical trajectories to obtain $\mathbb{P}_{ss}$ in principle. However since we need only one more BC, we only need to analyse a single trajectory. 
%for the ODE for $w(r)$. 
For this, we 
note that a trimer of three ABPs (labelled $1,2,3$) in an equilateral triangle
configuration (i.e. all separated by the zero-force radius $r_0 = 2^{1/6}+O(f_P)$)
is meta-stable \textit{if} their respective self-propulsion forces are directed to
the centre of the trimer. This implies that the relative velocities
of all three particles should be zero, i.e. $\bm{V}_{ij}=\bm{V}_{j}-\bm{V}_{i}$, then $\bm{V}_{12}=\bm{V}_{13}=\bm{V}_{23}=\bm{0}$ leading to \sc{the boundary condition
\begin{equation}\label{eq:wBC}
  2^{1/6}w'(2^{1/6})+w(2^{1/6}) = -1/3.
\end{equation}}
%\fx{\st{BC
%\begin{math}\label{eq:wBC}
% $2^{1/6}w'(2^{1/6})+w(2^{1/6}) = -\frac{1}{3}$
 %\end{math}
%}}
.}

While based on sound physical principles, the 
approximations we have applied are mathematically uncontrolled, and hence must be checked. We do this by empirical comparison with direct numerical simulations below.

We can now
calculate macroscopic properties using steady-state distribution, $\mathbb{P}_{ss}$. The expectation value of any observable $\mathcal{O}$ is 
\begin{equation}\label{eq:genericobservable}
  \big<\mathcal{O}\big> =\frac{1}{Z} \int \bigg(\prod_{i=1}^Nd^2r_id\theta_i\bigg)
  \mathcal{O}(\vec{X})e^{-\frac{1}{2}\sum_{j,k\neq j}h_{jk}}
\end{equation}
where \begin{math}\label{eq:hij}
  h_{ij} = \mathcal{V}(r_{ij})
  +f_Pw(r_{ij})\bm{r}_{ij}\cdot\bm{u}_{ij} \; , \; 
\end{math}
and 
the normalisation constant (the ``partition function'')
is \sc{
\begin{equation}\label{eq:partition}
  Z = \int \bigg(\prod_{i=1}^Nd^2r_id\theta_i\bigg)
  e^{-\frac{1}{2}\sum_{j,k\neq j}h_{jk}}.
\end{equation}}
%\fx{\st{
%\begin{math}\label{eq:partition}
%  $Z = \int \bigg(\prod_{i=1}^Nd^2r_id\theta_i\bigg) e^{-\frac{1}{2}\sum_{j,k\neq j}h_{jk}}.$
  %\end{math}
%}}
This is our first main result. 
%Importantly, no fitting parameters are introduced in deriving this.
%
Furthermore, one can explicitly integrate out
orientational degrees of freedom  in $\mathbb{P}_{ss}$ \cite{appendix}
to arrive at the marginal distribution
$\mathcal{Q}_{ss}(\bm{r}_1,\ldots,\bm{r}_N)\sim e^{-U_{\mathrm{eff}}(\{\bm{r}_i\})}$ with
an effective potential dependent only on particle positions
\footnote{The ability to easily integrate out orientational degrees of freedom limits
  $h_{ij}$ to being linear in $\bm{u}_{i}$. Thus, if a more generic form of
  $h_{ij}$ were to be derived which was still linear in $\bm{u}_{i}$, one could
  still obtain an effective potential through this integration procedure.},
\begin{equation}\label{eq:effectivepotential}
  \begin{split}
    U_{\mathrm{eff}} =& \frac{1}{2}\sum_{i,j\neq i}\big(\mathcal{V}(r_{ij})
    -\frac{f_P^2}{2}w^2(r_{ij})r_{ij}^2\big)\\
    &-\frac{f_P^2}{4}\sum_{\substack{i,j\neq i \\ k\neq i, k\neq j}}
    w(r_{ij})w(r_{ik})\bm{r}_{ij}\cdot\bm{r}_{ik}+O(f_P^3).
  \end{split}
\end{equation}
This effective potential constitutes the second main result of this letter.
The self-propulsion $f_P\neq 0$ introduces a minimum in the effective interaction~\cite{appendix}. 
%eqn. (\ref{eq:effectivepotential}) 
It also
demonstrates the importance of three-body and higher-body terms, which arise due to the coupling between position and orientation degrees of freedom.
We note that the three-body effective interactions have also been observed in three dimensional simulations of ABPs~\cite{Turci2021} and an effective
potential of the Active-Ornstein-Uhlenbeck model \cite{PhysRevLett.117.038103}.
To obtain the average of a macroscopic
observable  where $\mathcal{O}$ is independent
of orientational degrees of freedom, one may do so using
$e^{-U_{\mathrm{eff}}}$ as the probability measure.

Next we use eqn. (\ref{eq:genericobservable}), to calculate the interacting
swim pressure of ABPs~\cite{C5SM01412C,Solon2015}. We first remind the reader that the equation of state of a 2D fluid with temperature $T$ at equilibrium whose particles interact via a pair potential, $\mathcal{V}(r)$ is \begin{math} p^{} = k_B T \rho + p^{}_v \; ,\end{math} with the virial interaction pressure, \begin{math}\; p_v = -\frac14\rho^2 \int_0^\infty 2 \pi r^2 \mathcal{V}'(r) g^{(2)}(r) dr\end{math}~\cite{AT87} 
where $g^{(2)}(r)$ is the radial distribution function. We aim to obtain an equivalent for ABPs.
Starting from the Kirkwood definition of the stress tensor~\cite{DoiBook86}, the total pressure for ABPs may split up as
\begin{math}\label{eq:press}
  p = \rho \bigg(1+\frac{f_P^2}{2D_r}\bigg) +
  p_v + p_s^{\mathrm{int}},
\end{math}
with the first term corresponding to the active ideal gas
pressure (since $D_t\gamma_t=1$ in our units), the second is the virial interaction
pressure $p_v $ \cite{appendix} (present in both passive and
active systems), and the last term is the interacting swim pressure $p_s^{\mathrm{int}}$
(which is non-zero only in the active case). The microscopic definition
of $p_s^{\mathrm{int}}$ for the  WCA potential $\mathcal{V}(r_{ij})$ is
\cite{C5SM01412C}
\begin{equation}\label{eq:intswimpress}
  p_s^{\mathrm{int}} = -\frac{f_p}{4 D_r A}\sum_{i,j\neq i}
  \bigg<\bm{u}_{ij}\cdot\bm{r}_{ij}\frac{\mathcal{V}'(r_{ij})}{r_{ij}}\bigg> \; .
\end{equation}
Using eqn. (\ref{eq:genericobservable}), we
 arrive at the expression for the steady-state swim pressure
\begin{equation}\label{eq:intswimpress_calc}
  p_s^{\mathrm{int}} = \frac{2\pi f_P^2}{4 D_r}(\rho^2a_2(\rho) + \rho^3 a_3(\rho))
  + O(f_P^3)
\end{equation}
where \sc{
\begin{subequations}
  \begin{align}
    a_2(\rho) &\equiv
    \int  r^2w(r)\mathcal{V}'(r)g^{(2)}_0(r)dr, \\
    a_3(\rho) &\equiv
    \int \int r^2w(r) s\mathcal{V}'(s)G^{(3)}_0(r,s)  drds
  \end{align}
\end{subequations}}
%\fx{\st{
%  \begin{math}
%    $a_2(\rho) \equiv     \int  r^2w(r)\mathcal{V}'(r)g^{(2)}_0(r)dr$%  \end{math} , and  \begin{math}
%    $a_3(\rho) \equiv    \int \int r^2w(r) s\mathcal{V}'(s)G^{(3)}_0(r,s)  drds  ,$
%  \end{math}
%  }}
are  {density dependent} functions that are obtained from  
the passive ($f_P=0$) two-body: $g_0^{(2)}(r)$
and three-body: $G_0^{(3)}(r,s)$ radial distribution functions  
%respectively 
\cite{appendix} 
%\fx{\st{with 
%\begin{math}
%  $G^{(3)}_0(r,s) \equiv \int_0^{2\pi}g_0^{(3)}(r,s,\theta_{rs}),   \cos\theta_{rs}  d\theta_{rs}$
%}}
%\end{math}
~\footnote{$g^{(3)}(r,s,\theta_{rs})$ is the triplet distribution
  function, which is related to the probability of having one particle at
  the origin, a second particle at radial distance $r$, and a third particle
  at radial distance $s$, with $\hat{\bm{r}}\cdot\hat{\bm{s}}=\cos\theta_{rs}$.}
  (see Figures \ref{fig:1}A(ii) and \ref{fig:1}A(iii)).
  This is our third main result.

\tl{In Figure \ref{fig:1}B, we compare eqn. (\ref{eq:intswimpress_calc}), (black solid line) to $p_s^{\mathrm{{int}}}$ directly measured from simulation.
We scale the theory by its value, $p_0^{\mathrm{th}}$
    at $\rho=0.01$ and the simulations by $p_0^{\mathrm{sim}} = {p_s^{\mathrm{int}} \over \rho^2 f_P^2}(f_P=0.25,\rho=0.01)$. With the scaling from eqn. (\ref{eq:intswimpress_calc}), the simulation data does indeed collapse onto a a single
curve in agreement with our calculation (the collapse is less noisy for higher densities due to more frequent collisions). We find $p_0^{\mathrm{sim}}/p_0^{\mathrm{th}} \approx 1.2$.
%(via eqn. (\ref{eq:intswimpress}). 
%We appropriately scale $p_s^{\mathrm{int}}$
We present simulation results for four different active force values, $f_P=0.25,0.50,1.0,2.0$,
as a function of density.  
We expect the theory to begin to deviate from simulations for $f_P > 1$.
The theoretical prediction of \tl{cubic} dependence in density is surprising as both $a_2 (\rho)$ and $\rho a_3(\rho)$ are density
dependent and so any higher dependence on $\rho$ must cancel. In fact, keeping only pair correlations gives the red dashed line which shows completely different behaviour. Clearly,
three-body correlations control the dependence of
$p_s^{\mathrm{int}}$ on density.
Finally at large $\rho$ the theory curve diverges. We expect this is due to higher-body correlations which are in principle calculable 
%(but cumbersome) 
in our framework.} 
%The black solid line in \ref{fig:1} is from eqn. (\ref{eq:intswimpress_calc}).
%The passive radial distribution functions $g^{(2)}_0$ and $G^{(3)}_0$ were evaluated numerically to obtain the theoretical curve. 

% The theorectical underestimation of the magnitude of swim pressure 
 %(by factor $\sim 1.2$) 
% in the theory is probably due to approximations in setting the BCs for $w$. 
% In particular the approximation of setting the cut-off at which $w(r)=0$, $r_c \rightarrow \infty$ which is not compatible with periodic BCs. 
 %{\color{red}{Setting $r_c$ to something like two particle diameters seems reasonable. But this is a bit ad-hoc.}}

\begin{figure*}[h]
  \includegraphics[width=\textwidth]{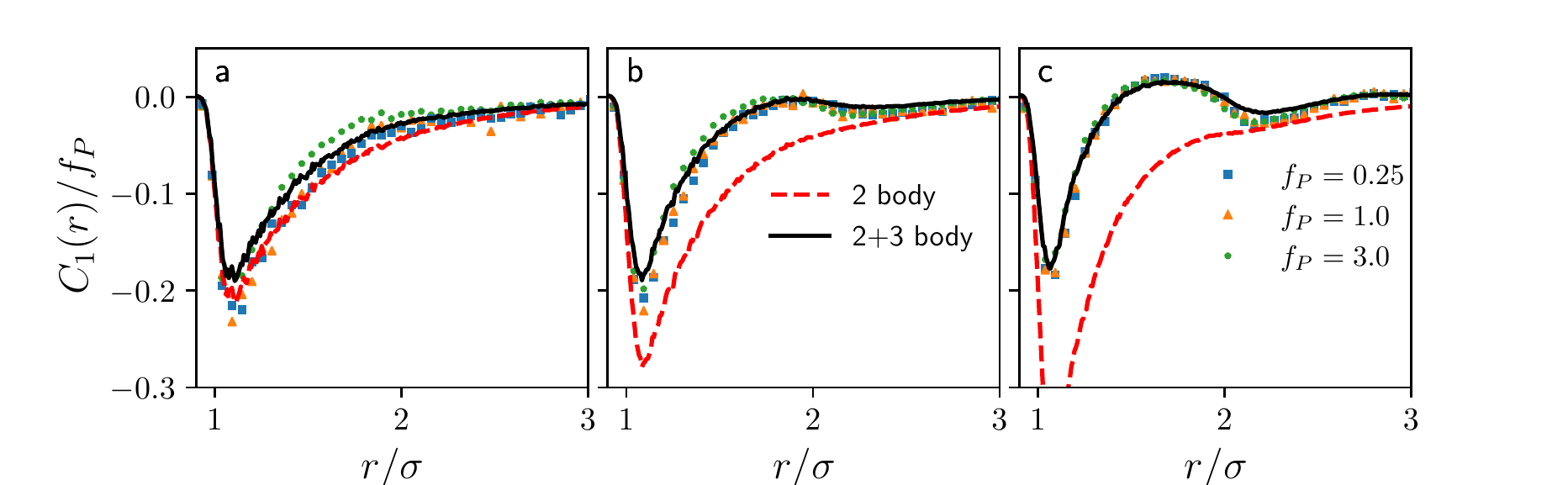}
  \caption{Pairwise orientational correlation function $C_1(r)$ scaled by
    self-propulsion $f_P$, vs interparticle distance $r$ \tl{for number densities
    (a) $\rho=0.1$, (b) $\rho=0.3$, and (c) $\rho=0.5$}. Filled markers correspond
    to simulation data. Pink squares, green triangles, and blue stars have
    self-propulsion values of $f_P=0.25$, $f_P=1.0$, and $f_P=3.0$, respectively.
   \label{fig:2}}
\end{figure*}

Finally, 
%For a second comparison between our theory and experiment, 
we compute the
local correlation function defined as
\begin{equation}\label{eq:C_1}
  C_1(r)\equiv
  \frac{1}{A}\bigg<\sum_{i,j\neq i}\bm{u}_{ij}\cdot\hat{\bm{r}}_{ij}
  \delta^2(\bm{r}_{ij}-\bm{r})\bigg> \;.
\end{equation}
Physically, this quantity represents angle-averaged correlations between
inter-particle orientation and inter-particle displacement, for a given
spacing $r$ between pairs of particles. 
\tl{It is interesting to measure this
function as it probes local structure of the ABPs, and so gives a more
detailed picture of whether our calculation captures the mesoscale structure as well as the thermodynamic behaviour of the system.}  We find $C_1$ also depends on two-body and three-body terms,
\begin{equation}\label{eq:C_1_calc}
  \frac{C_1(r)}{f_P^2\rho^2} =  -rw(r)g^{(2)}_0(r)
  - \rho \int s^2 w(s) G^{(3)}_0(r,s)  ds + \ldots 
\end{equation}
In Figure \ref{fig:2}, we measure $C_1(r)$ in simulation and compare our results
to those predicted by eqn. (\ref{eq:C_1_calc}). Again we find that with two-body terms only that the theoretical calculation does not agree with simulations and only compares well (for all $r$) on inclusion of the three-body terms (without fitting parameters).  
In conclusion, we have presented the first ab-initio calculation of the many-body steady-state probability density of states of Active Brownian Particles. This is based on a sequence of approximations that we check by numerical simulations.  \tl{It is promising that despite the approximations the calculation has managed to capture
the behaviour of  a number of local and global observables of the system without any free parameters}. 
 This indicates that these approximations are founded on sound physical principles and  
capture the essential features of the non-equilibrium steady-state of ABPs
at low $f_P$. It also suggests that they can form the starting point for a more detailed theory of these kinds of system 
%with fewer approximations 
that can be systematically improved by tightening the approximations and the assumptions behind them. 
%We anticipate that with a slightly more careful approach, one
%could obtain a perturbatively exact functional form for $\mathbb{P}_{ss}$
%and so rigorously map static ABP properties to an equilibrium calculation.

The computational resources of the University of Bristol Advanced Computing Research Centre, and the BrisSynBio HPC facility are gratefully acknowledged. TL acknowledges support of Leverhulme Trust Research Project Grant RPG-2016-147 and Bris- SynBio, a BBSRC/EPSRC Synthetic Biology Research Center (BB/L01386X/1).

\begin{comment}  
\appendix

\section{Correlation function calculations.\label{appsec:corr}}

\subfile{appendices/correlationfunctions.tex}

\section{Equivalence of two correlation functions.\label{appsec:equivalence}}

\subfile{appendices/equivalenceudotr.tex}

\section{Pressure of ABPs.\label{appsec:press}}

\subfile{appendices/pressure.tex}
\end{comment}

\bibliographystyle{h-physrev}
\bibliography{bib,library}

\end{document}